# Emergence of the SARS-CoV-2 furin cleavage site in humans is constrained to late 2019: implications for COVID-19 origins

Steven E Massey

Biology Dept, University of Puerto Rico at Rio Piedras, San Juan, PR 00901, USA

Email: steven.massey@upr.edu

Keywords: SARS-CoV-2, furin cleavage site, waiting time, D614G, compensatory mutation

**Abstract**

SARS-CoV-2 possesses a furin cleavage site (FCS) insertion in its spike protein, that is not found in other sarbecoviruses, and is responsible for its transmissibility in humans. The origin of the FCS insertion is unclear, and has been a point of debate regarding whether it arose naturally, or represents an artificial insertion. The FCS is destabilizing to spike protein, and as a consequence a compensatory mutation, D614G, rapidly emerged during the early pandemic and swept to fixation. Using a range of empirically determined parameters, a waiting time calculation can be used to determine the time interval between the appearance of the FCS and generation of the compensating D614G mutation. The waiting time is estimated to range from 37 to 65 days. Given the earliest detection of D614G on 1 January 2020, the latest dates for the appearance of the FCS range from October 28 to November 25 2019. This timeline is consistent with dates generated from phylogenetic considerations for the emergence of SARS-CoV-2 in late 2019, indicating that FCS appearance and virus emergence correspond. The waiting time approach enables discrimination between various scenarios of SARS-CoV-2 emergence, in particular cryptic circulation of a FCS-containing progenitor in the human population prior to late 2019 is unlikely, while prolonged circulation of a FCS-containing progenitor in bats appears inconsistent with absence of the compensatory mutation.

## Introduction

The COVID-19 pandemic was catalyzed by a short 4 amino acid sequence (PRRA) inserted into the S1/S2 boundary of the spike protein of SARS-CoV-2, which conferred a furin cleavage site (FCS) (Coutard et al. 2020). The FCS is cleaved by the endogenous protease furin, a process which separates the S1 and S2 spike protein subunits, altering the structure of the spike protein so that its receptor binding domain (RBD) can more readily adopt an open 'RBD-up' conformation, facilitating binding to the ACE2 receptor (Wrobel et al. 2020). The presence of the FCS substantially increases transmissibility of SARS-CoV-2 (Peacock et al. 2021), however a functional FCS at the S1/S2 junction is absent from all other sarbecoviruses, suggesting it is either maladaptive or selectively neutral in bats. Consequently, determining how the FCS was acquired is of great interest.

While the FCS insertion is adaptive, conferring a selective benefit to SARS-CoV-2, it is also structurally destabilizing to spike protein when cleaved (Wrobel et al. 2020), relaxing the trimer and weakening inter-trimer bonds (Barrett et al. 2021) and encouraging premature S1 shedding (Shoemaker et al. 2026). The D614G amino acid substitution is a compensatory mutation that arose early in the pandemic before sweeping rapidly to fixation (Korber et al. 2020). D614G ameliorates the destabilizing effects of FCS cleavage, by reducing premature S1 shedding, increasing the proportion of intact, functional trimers and shifting the conformational equilibrium to a more open (RBD-up) state; all of these effects facilitate receptor binding (Koenig and Schmidt 2021). Consistent with these observations, D614G results in increased infectivity in human cell lines of virus (Hou et al. 2020) (Zhou et al. 2021) (Plante et al. 2021), virus like particles (Yurkovetskiy et al. 2020) and pseudotyped viruses (Q. Li et al. 2020) (Zhang et al. 2020) (Korber et al. 2020). Likewise, D614G results in increased transmissibility in humanized mice and Syrian hamsters (Hou et al. 2020) and ferrets (Zhou et al. 2021). D614 is conserved in 44 sarbecovirus genomes (Jungreis et al. 2021), emphasising the atypical role of the radical D614G substitution in stabilizing the unique FCS insert in SARS-CoV-2 spike protein.

The evolutionary trajectory of the D614G substitution can help inform the timing of the FCS insertion given that once the FCS had been inserted then a selective pressure for the D614G substitution would immediately have arisen (Quay 2024). Given the high mutation rate of SARS-CoV-2 and its rapid growth, this means that the appearance of the substitution was predictable. Consequently, a waiting time calculation can be used to estimate when the FCS was inserted, given the earliest detection date of D614G, knowledge of its selection coefficient, and other characteristics of SARS-CoV-2 growth and spread. Using this approach, the latest timing of the insertion is dated from late October to late November 2019 and reveals insights into the origin of the pandemic.

## Methods

### *D614G prevalence analysis*

The Open LAPIS API (Chen et al. 2023) was used to query 295,095 SARS-CoV-2 sequences sampled from December 1 2019 to December 1 2020 for the presence of the A23403G single nucleotide variant (SNV), which produces the D614G amino acid substitution. Open LAPIS draws primarily from the National Center for Biotechnology Information (NCBI), along with the European Nucleotide Archive (ENA) and DNA Data Bank of Japan (DDBJ), collated by the International Nucleotide Sequence Database Collaboration (INSDC) (Karsch-Mizrachi et al. 2018).

A logistic curve was fitted to the increase in daily D614G prevalence values using the R nls package (Supplementary Material). The model used is the standard logistic (sigmoid) function:

$$p(t) = \frac{1}{1+e^{-k(t-t_0)}}$$

where *p(t)* is D614G prevalence at time *t*, *k* is the growth rate of D614G prevalence per day, and $t_0$ is the inflection point (day of 50% prevalence).

### *Waiting time model for the D614G mutation*

A stochastic waiting time model was used to calculate the expected time to the first successful compensatory mutation after the acquisition of the FCS. The model is derived from a non-homogenous Poisson process for the establishment of a beneficial mutation in an exponentially growing population:

$$T = \frac{1}{r} ln(1 + \frac{1}{\mu_s \cdot P_{est}})$$

where $T$ is the mean waiting time, $r$ is the epidemic growth rate per day, $\mu_s$ is the per-site mutation rate for the A→G transition, and $P_{est}$ is the probability of establishment (Haldane's survival probability). The derivation of the equation is described in the Supplementary Material.

$P_{est}$ can be approximated to *2s*, where *s* is the selection coefficient, when the population is large and *s* is weak (Haldane 1927). However, the early stages of the pandemic are expected to be marked by stochasticity such as superspreader events and lineage extinctions. This effect can be captured by use of a dispersion factor, K. The smaller the value of K, the higher the effect of stochastic events (overdispersion) and the lower the likelihood of a beneficial mutation establishing itself.

To account for overdispersion, the probability of establishment can be expressed as $P_{est} \approx \frac{2s}{\sigma^2}$, where $\sigma^2$ is the variance in the offspring distribution (Patwa and Wahl 2008). Assuming a negative binomial offspring distribution, the variance for a stable population (where the reproduction number, $R \approx 1$) can be expressed as $\sigma^2 \approx 1 + \frac{1}{\kappa}$ (Lloyd-Smith et al. 2005). Substituting this variance yields the *K*-adjusted probability of establishment:

$$P_{est} \approx \frac{2s}{1+\frac{1}{\kappa}}$$

However, in the early pandemic the population was growing. In this scenario, the *K-a*djusted probability of establishment can be expressed as :

$$P_{est} = 1 - P_{ext}$$

where $P_{ext}$ is the probability of extinction, which in turn can be expressed as (Czuppon et al. 2021):

$$z = \left( \frac{\frac{\kappa}{\kappa + R_{mut}}}{1 - \left(1 - \frac{\kappa}{\kappa + R_{mut}}\right) z} \right)^{\kappa}$$

where $z$ is the probability of extinction and $R_{mut}$ represents the reproduction number $R$ adjusted for the increase in fitness due to the D614G mutation, reflected in the selection coefficient $s$. This can be expressed as $R_{mut} = (1 + s)R$. The adaptive benefit of the D614G mutation is reflected in an increased reproduction number. The equation is solved for the smallest positive fixed point value of $z$ (Czuppon et al. 2021). This was done numerically using the R uniroot package (Supplementary Material).

The value of $s$ for the D614G mutation has been estimated as ranging from 0.10 (Volz et al. 2021) to 0.31 (Leung et al. 2021). $R$ in the early pandemic has been estimated as ranging from 2.2 (using the estimated number of infections on 18 Jan 2020, (Riou and Althaus 2020)) to 3.6 (using the number of cases in China from 10-24 Jan 2020, (Zhao et al. 2020)). These estimates were generated using data from infected individuals with a likely low prevalence of D614G. Later estimates of $R$ should derive from infected individuals with a higher prevalence of D614G, capturing the increased fitness of D614G mutants. However, they may also be (negatively) affected by the onset of public health measures. Consequently, these early estimates of $R$ can be used in conjunction with $s$ to generate $R_{mut}$, using the equation $R_{mut} = (1 + s)R$.

Other empirically determined parameters used for the waiting time analysis are as follows. The per transmission mutation rate for SARS-CoV-2 has been calculated as 0.7-1.0 nucleotide mutations per transmission (Murray et al. 2024). The A→G transition rate in the SARS-CoV-2 genome has been calculated as 40-70 % of the total average mutation rate (Yi et al. 2021). To calculate the per-site mutation rate for the A→G transition ($\mu_s$), it is necessary to divide the per transmission mutation rate by the number of sites in the SARS-CoV-2 reference genome (29,903 nucleotides). Then $\mu_s$ is calculated taking into account the A→G transition rate. The growth rate from the early pandemic (January 15–30, 2020 in Wuhan) has been estimated as ranging from 0.21 day$^{-1}$ to 0.30 day$^{-1}$ (Sanche et al. 2020). Estimates of the dispersion factor *K* in the early pandemic range from 0.10 (Endo et al. 2020) to 0.54 (Riou and Althaus 2020).

## Results

### *D614G prevalence analysis*

A logistic curve was fitted to daily D614G prevalence values generated from 295,095 INSDC sequences. The curve produces an inflection point of 6 March 2020 and a growth value of $k$ = 0.0495 day$^{-1}$. Using $\frac{ln2}{k}$, the log-odds doubling time of D614G prevalence was calculated as 14 days, emphasizing the speed with which it spread to fixation.

Prior to the inflection point, sample sizes are relatively small and sparse, which means that while in principle the date of emergence for this D614G substitution can be estimated from curve fitting, it has low confidence. This is reflected in the residuals corresponding  to dates prior to 6 March 2020 (Supplementary Figure 1).

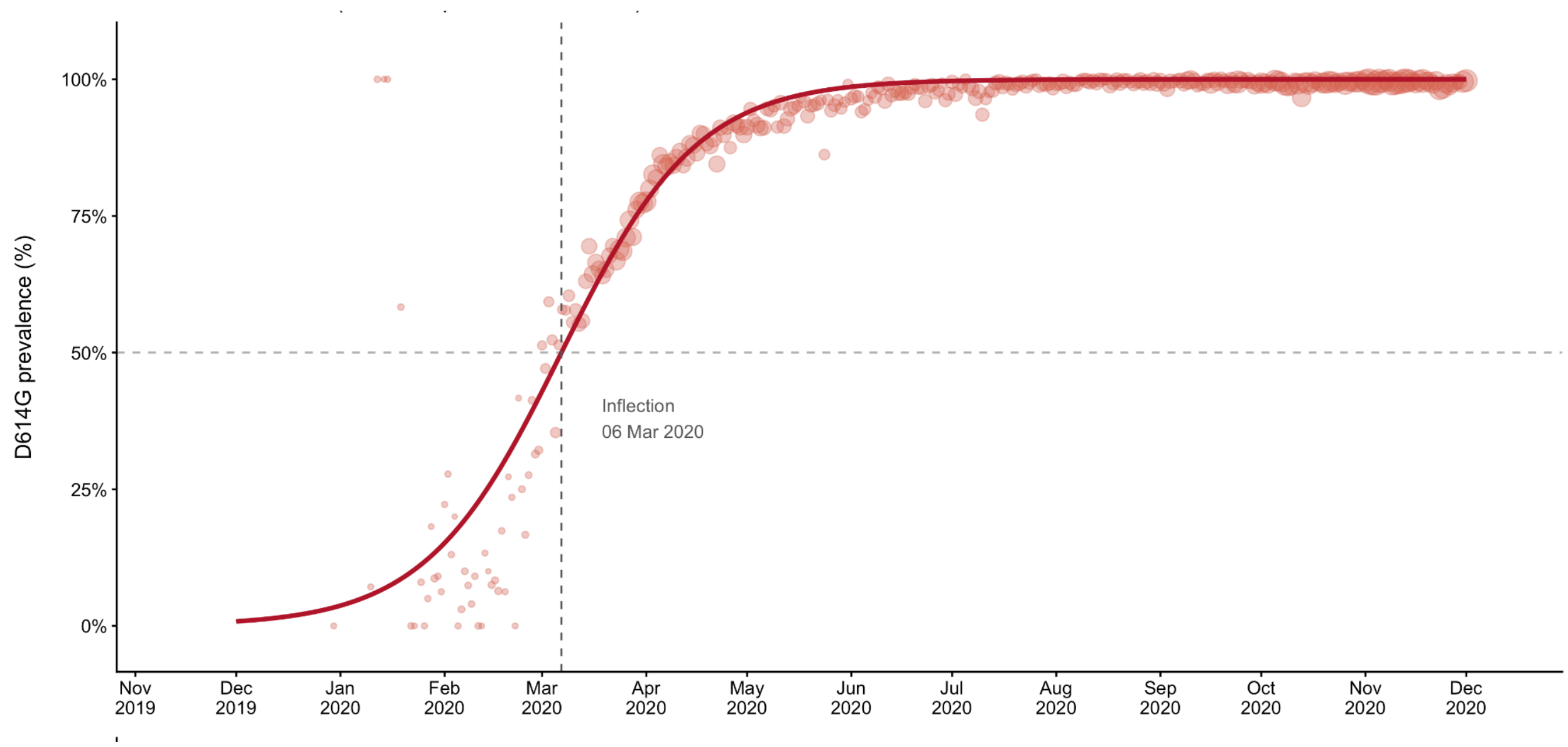

**Figure 1** D614G prevalence from 1 Dec 2019 to 1 Dec 2020
Each day's prevalence was weighted by the square root of the total sequences collected that day. Days with less than 10 sequences were excluded from the analysis, and the analysis was restricted to the sweep phase, for frequency values between 0.1 % and 99.9 %. Due to these filters 299 days were used for the curve fitting. The size of the data points are proportional to the number of samples collected that day.

*Waiting time calculation*

| Scenario | r (day$^{-1}$) | $\mu_{total}$ | f_A→G | s | R | κ | $R_{mut}$ | $\mu_s$ (×10$^{-6}$) | z | $P_{est}$ | T (days) |
|---|---|---|---|---|---|---|---|---|---|---|---|
| **Minimum Waiting Time** | 0.300 | 1.00 | 0.70 | 0.310 | 3.6 | 0.54 | 4.7160 | 23.41 | 0.3617 | 0.6383 | **37.04** |
| **Maximum Waiting Time** | 0.210 | 0.70 | 0.40 | 0.100 | 2.2 | 0.10 | 2.4200 | 9.36 | 0.8649 | 0.1351 | **64.67** |
| **Mean Estimate** | 0.255 | 0.85 | 0.55 | 0.205 | 2.9 | 0.32 | 3.4945 | 15.63 | 0.5748 | 0.4252 | **46.75** |
| Low Growth Rate Only | 0.210 | 0.85 | 0.55 | 0.205 | 2.9 | 0.32 | 3.4945 | 15.63 | 0.5748 | 0.4252 | **56.77** |
| High Growth Rate Only | 0.300 | 0.85 | 0.55 | 0.205 | 2.9 | 0.32 | 3.4945 | 15.63 | 0.5748 | 0.4252 | **39.74** |

| | | | | | | | | | | |
|---|---|---|---|---|---|---|---|---|---|---|
| Low S Only | 0.255 | 0.85 | 0.55 | 0.100 | 2.9 | 0.32 | 3.1900 | 15.63 | 0.5967 | 0.4033 | **46.96** |
| High S Only | 0.255 | 0.85 | 0.55 | 0.310 | 2.9 | 0.32 | 3.7990 | 15.63 | 0.5556 | 0.4444 | **46.58** |
| Low R Only | 0.255 | 0.85 | 0.55 | 0.205 | 2.2 | 0.32 | 2.6510 | 15.63 | 0.6446 | 0.3554 | **47.45** |
| High R Only | 0.255 | 0.85 | 0.55 | 0.205 | 3.6 | 0.32 | 4.3380 | 15.63 | 0.5267 | 0.4733 | **46.33** |
| Low KAPPA Only | 0.255 | 0.85 | 0.55 | 0.205 | 2.9 | 0.10 | 3.4945 | 15.63 | 0.8197 | 0.1803 | **50.12** |
| High KAPPA Only | 0.255 | 0.85 | 0.55 | 0.205 | 2.9 | 0.54 | 3.4945 | 15.63 | 0.4361 | 0.5639 | **45.64** |
| Low Mutation Rate Only | 0.255 | 0.70 | 0.40 | 0.205 | 2.9 | 0.32 | 3.4945 | 9.36 | 0.5748 | 0.4252 | **48.76** |
| High Mutation Rate Only | 0.255 | 1.00 | 0.70 | 0.205 | 2.9 | 0.32 | 3.4945 | 23.41 | 0.5748 | 0.4252 | **45.17** |

**Table 1** Waiting Time estimates generated under varying parameter values

The values in the table were generated using the waiting time calculation described in Methods, and the corresponding parameter values.

The minimum waiting time is 37 days, while the maximum is 65 days, with an average of 47 days (Table 1). The sensitivity analysis in Table 1 indicates that variation in the growth rate has the largest effect on the estimated waiting time. The selection coefficient and reproduction number only have minimal effects on the estimated waiting time. The earliest D614G mutation was sampled on 1 January 2020 in Argentina (EPI_ISL_4405694) (Hu et al. 2023), this date can be used as an anchor for the waiting time calculation. Using this, dates ranging from October 28 to November 25 2019 are generated for the appearance of the FCS in infected humans.

## Discussion

Comparison with spike sequences of related sarbecoviruses indicates that the FCS was acquired as a single insertion, as there are no intermediate sequences (Chan and Zhan 2022). Once the FCS(+) progenitor was present in humans, our results indicate the clock began ticking for the acquisition and establishment of the compensatory D614G mutation. This has a deterministic element, given its dependence on the number of transmissions, which grew exponentially in the early pandemic.

D614G arose at least five times independently in the early pandemic (van Dorp, Richard, et al. 2020). The origin of the specific D614G mutation that spread rapidly to fixation in 2020 has been dated to 18 January 2020, in the B.1 lineage, which expanded from Europe (Isabel et al. 2020). The plot demonstrates the speed of fixation of the mutation, reflecting a strong selective pressure. A caveat is that the SNV C14408T, which produces the P322L amino acid substitution in the nsp12 RNA-dependent RNA polymerase (RdRp) protein, also has an adaptive benefit and is linked to D614G in the haplotype that swept to fixation (Ilmjärv et al. 2021).

According to the sensitivity analysis, the most important factor for the speed of establishment of D614G is the growth rate. The growth rate values used in the analysis date from early 2020, there are no data from late 2019. Variation in growth rate in the earliest stage of the pandemic in 2019 would be expected to affect the waiting time estimates therefore, no matter which model is used. However, this seems unlikely as the growth rates used here were calculated from case reports (Sanche et al. 2020), the disease characteristics are unlikely to have differed substantially earlier in the outbreak.

The dates of October 28 to November 25 2019 are the latest dates for the appearance of the FCS, as the D614G mutation likely pre-dated its sampling on 1 January 2020 by a number of days. Estimates for the time to the most recent common ancestor (tMRC) range from mid November (Nie et al. 2020), late November (X. Li et al. 2020) (Benvenuto et al. 2020), early October to early December (van Dorp, Acman, et al. 2020), late October to mid November (Kumar et al. 2021) and August to early October (Samson et al. 2024). These are largely consistent with the waiting time calculation presented here, indicating that the FCS was present at the very start of the pandemic.

A feature apparent from the waiting time analysis is the speed with which the D614G mutation is expected to establish after destabilizing FCS cleavage becomes manifest.

The implications of these observations for the origin of COVID-19 are discussed in a variety of possible scenarios of emergence in the human population, as follows.

*1. Cryptic circulation in humans of SARS-CoV-2 progenitor lacking a FCS, followed by acquisition of the FCS*

In this scenario, a FCS(-) SARS-CoV-2 progenitor circulated cryptically in humans, before insertion of the FCS, which catalyzed the spread of the virus, initiating the pandemic. Our data indicate the timing of the acquisition of the FCS as October to November 2019. However, FCS(-) SARS-CoV-2 would likely have had limited transmissibility in humans, as contact transmission of a deletion mutant does not occur in ferrets (Peacock et al. 2021). Limited transmissibility in humans is also consistent with the vanishingly low number of sequences that have deletions in the FCS in GISAID (Peacock et al. 2021) (Nagy et al. 2021).

A related scenario is cryptic circulation of a FCS(+) SARS-CoV-2 progenitor prior to the start of the pandemic. The contemporaneity of the waiting time calculation with clock estimates of SARS-CoV-2's emergence are inconsistent with this scenario.

*2. Zoonosis from an intermediate or reservoir host*

The waiting time calculation indicates that the FCS appeared in humans at a time contemporaneous to the emergence of SARS-CoV-2 in the human population. This suggests that a FCS(+) SARS-CoV-2 progenitor may not have spent much time in an intermediate or reservoir host. Given the failure to identify an intermediate host that efficiently transmits FCS(+) SARS-CoV-2, identifying an intermediate host that transmits a FCS(-) progenitor is problematic. Extended presence of FCS(+) SARS-CoV-2 in a host animal can be predicted to lead to the emergence of the D614G mutation, given its compensatory nature, assuming that FCS cleavage was beneficial in the host animal.

*3. Introduction into humans from a research sample*

If FCS(+) SARS-CoV-2 was present in a research sample from a bat or other animal, that then led to accidental infection of a researcher, its presence in the sampled animal would imply circulation in that animal population. This in turn implies that FCS cleavage was beneficial in that animal. Under these circumstances, the D614G substitution can be expected to arise, given its compensatory benefit countering the destabilizing effects of cleavage. Absence of D614G in early SARS-CoV-2 sequences, suggests this scenario is disfavored, however this requires further investigation.

*4. Serial passaging in a lab animal before lab escape*

Serial passaging of coronaviruses in lab animals typically involves a limited number of passages, with each passage representing a bottleneck for the virus. The bottleneck reduces the strength of selection, which would increase the time for a beneficial mutation to become fixed. Consequently, a FCS(+) SARS-CoV-2 could feasibly have been passaged a limited number of times without picking up D614G, prior to lab escape.

*5. Serial passaging in cell culture before lab escape*

Serial passaging in cell cultures involves taking a small aliquot after each virus amplification step, which is then used to inoculate a new culture. This results in a bottlenecking effect which reduces the effective population size. This is expected to weaken the strength of selection, slowing the acquisition of a compensatory mutation. The lack of D614G appears compatible with cell culture serial passaging of FCS(+) SARS-CoV-2, prior to lab escape therefore, however this requires further investigation.

*6. Direct infection of lab personnel from infectious clone DNA*

This scenario involves improper handling of a complete circular infectious clone DNA construct that harbored the FCS(+) SARS-CoV-2 genome. Given its circularity, such an infectious clone would be in an transfective state (Wong et al. 2021). Exposure to such an infectious clone may lead to infection of a lab worker via direct contact, if not handled correctly. Lack of the compensatory D614G in early SARS-CoV-2 genomes is consistent with this scenario, as it would not have had a chance to arise prior to infection of the lab worker.

## Conclusion

The waiting time calculation presented here indicates that the D614G compensatory mutation is expected to arise rapidly in humans, within 1-2 months circulation of a FCS containing SARS-CoV-2 progenitor. Dates for the appearance of a FCS-containing progenitor in the human population range from late October to late November, and are roughly contemporaneous with clock based estimates. The dates for the insertion appear compatible with several lab escape scenarios, but inconsistent with prolonged circulation of a FCS(+) progenitor in bats, cryptic spread in humans or infection of a lab worker from a research sample. However, further investigation on this topic is warranted.

## Acknowledgements

The authors would like to thank members of the DRASTIC collective and to Michael Weissman (University of Illinois) for stimulating discussion on topics addressed in this work.

**Supplementary Material**

**Supplementary File 1** Daily D614G counts

`d614g_daily_counts.csv`

**Supplementary Figure 1** Residuals for the fitted logistic D614G prevalence curve

In order to test goodness of fit of the logistic curve, residuals for each day *i* were calculated as the difference between the observed prevalence of D614G and model-predicted prevalence, as follows: $Residual_i = (p_{observed,i} - p_{predicted,i})\ x\ 100$.

Residuals are expressed as percentage points (pp). The residuals generated from Figure 1 are mostly small (±5 pp) after April 2020 and show no strong systematic trend indicating that the logistic model captures the overall growth well. Larger and absent residuals between February–March 2020 reflect small sample sizes.

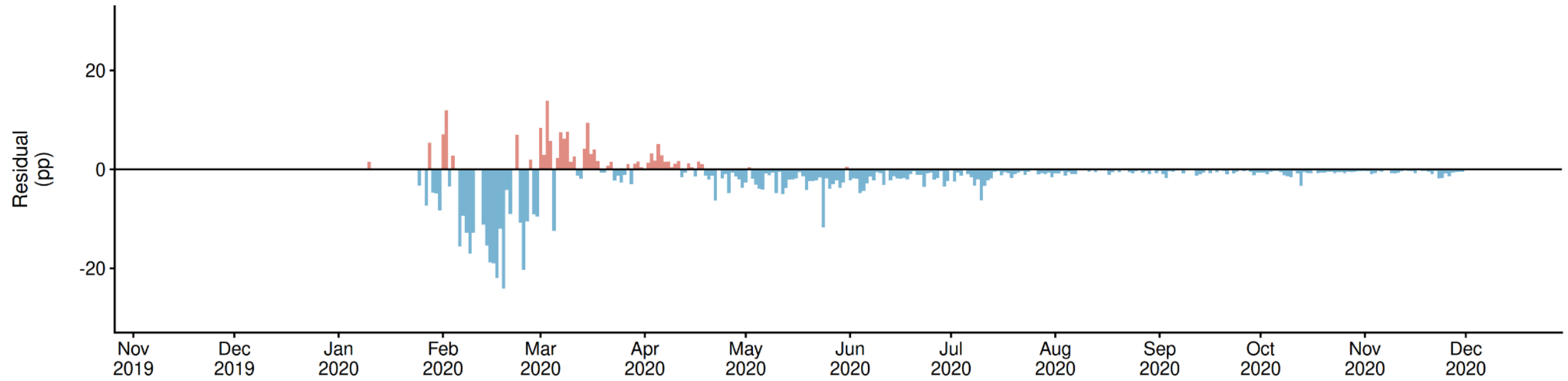

**Supplementary calculation**

*Derivation of the waiting time model*

The FCS-containing virus is considered to spread exponentially in the early stages of the pandemic. If the epidemic starts with 1 individual infected with FCS-containing SARS-CoV-2 at time $t = 0$, the cumulative number of transmission events (total infections) *C(T)* up to time *T* can be approximated to the exponential growth function:

$$C(T) \approx e^{rT} - 1$$

where *r* is the epidemic growth rate. The probability that a specific mutation (e.g., A→G at site 23403) occurs during a single transmission is the per-site mutation rate, $\mu_s$

The probability that the mutation survives and establishes a lasting lineage is $P_{est}$ (probability of establishment). Therefore, the probability that any single transmission event results in a successful, established mutation is $\mu_s \cdot P_{est}$

The expected number of successful mutations is $C(T) \cdot (\mu_s \cdot P_{est})$. Substituting the value of $C(T)$ from above gives:

$$Expected\ successful\ mutations = (e^{rT} - 1) \cdot \mu_s \cdot P_{est}$$

The waiting time *T* is the time when the number of successful mutations reaches 1, leading to the following:

$$(e^{rT} - 1) \cdot \mu_s \cdot P_{est} = 1$$

The equation is solved for *T*, giving the waiting time calculation:

$$T = \frac{1}{r} ln(1 + \frac{1}{\mu_s \cdot P_{est}})$$

**Analysis Command Lines**

Manus 1.6 Lite AI agent (https://manus.im) was used for the generation of Open LAPIS API and R command lines, these were visually checked, manually edited and manually implemented.

*Open LAPIS query*

```
# To download all sequences from 1 Dec 2019 to 1 Dec 2020
curl -s
"https://lapis.cov-spectrum.org/open/v2/sample/aggregated?dateFr
om=2019-12-01&dateTo=2020-12-01&fields=date" -o
total_counts.json

# To count the number of D614G mutants between the same dates
curl -s
"https://lapis.cov-spectrum.org/open/v2/sample/aggregated?dateFr
om=2019-12-01&dateTo=2020-12-01&nucleotideMutations=A23403G&fiel
ds=date" -o d614g_counts.json

# To count the total number of sequences (295095)
cat total_counts.json | jq '[.data[].count] | add'

# To validate the number of D614G mutants (276962)
cat d614g_counts.json | jq '[.data[].count] | add'

# Create a csv file:
echo "date,total,d614g,prevalence_pct" > d614g_daily_counts.csv
```

```
# Add the counts to the csv file:
jq -r -s '
  (.[0].data | map({(.date): .count}) | add) as $totals |
  (.[1].data | map({(.date): .count}) | add) as $d614g |
  ($totals | keys) + ($d614g | keys) | unique | sort |
  map(
    . as $date |
    ($totals[$date] // 0) as $t |
    ($d614g[$date] // 0) as $d |
    (if $t > 0 then (($d / $t * 10000) | round / 100) else ""
end) as $pct |
    [$date, $t, $d, $pct] | @csv
  ) | .[]
' total_counts.json d614g_counts.json >> d614g_daily_counts.csv
```

*Curve fitting R commands*

```
# install packages
install.packages("ggplot2")
install.packages("dplyr")
install.packages("scales")
install.packages("patchwork")
install.packages("lubridate")

# load packages
suppressPackageStartupMessages({
  library(ggplot2)
  library(dplyr)
  library(scales)
  library(patchwork)
  library(lubridate)
```

```
})

# Load data
df <- read.csv("d614g_daily_counts.csv")
df$date <- as.Date(df$date)

# Filter data for fitting (minimum 10 sequences, exclude 0% and
100% saturation)
df_fit <- df %>%
  filter(total >= 10, !is.na(prevalence_pct)) %>%
  mutate(
    prev_frac = d614g / total,
    t = as.numeric(date - as.Date("2019-12-01"))
  )

df_sweep <- df_fit %>%
  filter(prev_frac > 0.001, prev_frac < 0.999)

# Fit logistic model using nls, weighted by sqrt(total
sequences)
# p(t) = 1 / (1 + exp(-k * (t - t0)))
fit <- nls(
  prev_frac ~ 1 / (1 + exp(-k * (t - t0))),
  data = df_sweep,
  start = list(k = 0.05, t0 = 100),
  weights = sqrt(total)
)

params <- coef(fit)
k_fit <- params["k"]
t0_fit <- params["t0"]
```

```
inflection_date <- as.Date("2019-12-01") + t0_fit
logodds_doubling <- log(2) / k_fit

# Generate smooth curve data for plotting
t_plot <- seq(0, as.numeric(as.Date("2020-12-01") -
as.Date("2019-12-01")), length.out = 500)

curve_df <- data.frame(
  t = t_plot,
  date = as.Date("2019-12-01") + t_plot,
  prev_frac = 1 / (1 + exp(-k_fit * (t_plot - t0_fit)))
)

# Calculate residuals
df_sweep$pred <- 1 / (1 + exp(-k_fit * (df_sweep$t - t0_fit)))

df_sweep$residual <- (df_sweep$prev_frac - df_sweep$pred) * 100
```

*Calculation of z (probability of extinction)*

```
s     <- 0.205    # example selection coefficient
R     <- 2.9      # example basic reproduction number
kappa <- 0.32     # example dispersion factor

R_mut <- (1 + s) * R    # = 3.4945

# —— Extinction-probability fixed-point equation
———————————————————————————————
# Solve for the smallest positive z satisfying:
#
#          /        kappa / (kappa + R_mut)         \ ^kappa
#   z  =  | ---------------------------------------- |
```

```
#           \ 1 - (1 - kappa/(kappa+R_mut)) * z       /
#
# Rearranged as f(z) = 0:
#   f(z) = z * (1 - q * z)^kappa  -  p^kappa
# where p = kappa / (kappa + R_mut),  q = 1 - p

extinction_prob <- function(R_mut, kappa) {
  if (R_mut <= 1) return(1)  # certain extinction
  p <- kappa / (kappa + R_mut)
  q <- 1 - p
  f <- function(z) z * (1 - q * z)^kappa - p^kappa
  uniroot(f, interval = c(1e-9, 1 - 1e-9), tol = 1e-12)$root
}

z  <- extinction_prob(R_mut, kappa)
```

*Calculation of waiting time*

```
calculate_waiting_time <- function(r, mu_s, P_est) {
  # T = (1 / r) * ln(1 + 1 / (mu_s * P_est))
  T <- (1 / r) * log(1 + (1 / (mu_s * P_est)))
  return(T)
}

r_val <- 0.3          # Example epidemic growth rate per day
mu_s_val <- 23.41e-6  # Example per-site mutation rate
P_est_val <- 0.6383   # Example probability of establishment

mean_time <- calculate_waiting_time(r_val, mu_s_val, P_est_val)
```